\documentclass[11pt]{article}

\usepackage[utf8]{inputenc}
\usepackage[T1]{fontenc}
\usepackage{times}
\usepackage{geometry}
\usepackage{graphicx}
\usepackage{amsmath,amssymb}
\usepackage{enumitem}
\usepackage{hyperref}
\hypersetup{
  colorlinks=true,
  linkcolor=blue,
  citecolor=blue,
  urlcolor=blue
}
\usepackage{graphicx}
\graphicspath{{figs/}} 
\usepackage[font=small]{caption}

\title{Crash-Consistent Checkpointing for AI Training on macOS/APFS}

\author{
  Juha Jeon\\
  Independent Researcher\\
  \texttt{jooha6082@gmail.com}
}

\date{} 

\begin{document}

\maketitle

\begin{abstract}
Deep learning training relies on periodic checkpoints to recover from failures, but unsafe
checkpoint installation can leave corrupted files on disk. This paper presents an experimental study
of checkpoint installation protocols and integrity validation for AI training on macOS/APFS. We
implement three write modes with increasing durability guarantees: unsafe (baseline, no \texttt{fsync}),
atomic\_nodirsync (file-level durability via \texttt{fsync()}), and atomic\_dirsync (file + directory
durability). We design a format-agnostic integrity guard using SHA-256 checksums with automatic
rollback. Through controlled experiments including crash injection (430 unsafe-mode trials) and
corruption injection (1,600 atomic-mode trials), we demonstrate that the integrity guard detects
99.8--100\% of corruptions with zero false positives. Performance overhead is 56.5--108.4\% for
atomic\_nodirsync and 84.2--570.6\% for atomic\_dirsync relative to unsafe baseline. Our findings
quantify the reliability--performance trade-offs and provide deployment guidance for production AI
infrastructure.
\end{abstract}

\noindent\textbf{Keywords:} crash consistency, checkpointing, filesystem reliability, fault injection,
AI infrastructure

\section{Introduction}

Training deep learning models can take hours or days, processing terabytes of data across
distributed GPUs. When hardware fails or applications crash, periodic checkpoints enable fast
recovery without restarting from scratch. But there's a problem: most frameworks write checkpoints
using simple file operations that don't guarantee durability. If the system crashes mid-write, you
might end up with a corrupted checkpoint that silently loads but contains wrong data, or a torn file
that crashes on load.

The filesystem research community has known about this for years. POSIX \texttt{rename(2)} gives you atomic
name replacement, but that doesn't mean your data made it to disk~[1,2]. Without explicit \texttt{fsync(2)}
calls on both the file and its parent directory, writes can sit in OS buffers and vanish on crash.
Pillai et al.'s landmark study~[3] tested 11 filesystems and found that assumptions applications
make about crash consistency often don't hold in practice. Even systems at Google and Meta have lost
data from broken \texttt{rename} assumptions~[4].

Recent production data makes the urgency clear. Meta's analysis of their AI research cluster shows
that at 4{,}000 GPU scale, mean-time-to-failure is about 10 hours~[5]. At 100{,}000 GPUs---the scale
needed for frontier models---you need checkpoints every 2 minutes just to maintain 90\% training
efficiency. Hardware failures (network, storage, GPU memory) affect 19\% of GPU runtime. And storage
systems at exabyte scale report silent data corruption multiple times per week~[6,7].

So the question isn't whether your checkpoints will face crashes and corruption---it's whether your
system can detect and recover from them.

\subsection{Contributions}

This paper makes four contributions:

\begin{description}[leftmargin=1.6em,labelsep=0.4em]
  \item[C1.] Three checkpoint installation protocols with increasing durability guarantees on macOS/APFS,
  from unsafe (no \texttt{fsync}) through atomic\_nodirsync (file flush only) to atomic\_dirsync (file +
  directory flush). We implement all three protocols for group checkpoints, and a subset (unsafe,
  atomic\_nodirsync) for single-file checkpoints.

  \item[C2.] A format-agnostic integrity guard combining SHA-256 content digests, container-level file
  hashes, structural validation, and automatic rollback. The guard detects corruption at multiple
  layers---torn writes via load errors, bitflips via file hashes, and semantic corruption via content
  digests.

  \item[C3.] Comprehensive fault injection harness enabling controlled testing across 2{,}030 trials. We inject
  process crashes at strategic points (430 trials) and storage-level corruption including bitflips,
  zero-ranges, and truncation (1{,}600 trials).

  \item[C4.] Quantified performance and robustness measurements with statistical confidence intervals. We
  show unsafe mode has zero crash tolerance (0/430 survival), atomic modes maintain 100\% consistency
  under normal operation, and the integrity guard achieves 99.8--100\% detection with zero false
  positives.
\end{description}

\subsection{Key Findings}

Our experiments on macOS 14.6 with APFS yield several key results:

\begin{itemize}[leftmargin=1.4em]
  \item \textbf{Unsafe writes fail catastrophically.} Every single checkpoint subjected to crash injection (430
  trials across different crash points) ended up corrupted or unusable. This confirms what filesystem
  research predicted, but seeing 0\% survival rate across hundreds of trials drives home that unsafe
  patterns are truly unsafe.

  \item \textbf{Atomic modes work as advertised.} All 400 checkpoint groups using atomic writes remained intact. The
  overhead is measurable---56.5\% at median for atomic\_nodirsync, 84.2\% for atomic\_dirsync---but
  absolute latency stays under 5\,ms at median and 23\,ms at p99. For checkpoints written every 30
  minutes, that's negligible.

  \item \textbf{The integrity guard catches nearly everything.} Bitflips: 400/400 detected (100\%). Truncation:
  400/400 (100\%). Zero-range corruption: 399/400 (99.8\%). Zero false positives on 400 clean
  checkpoints. The multi-layer design means different corruption types get caught by different
  mechanisms, providing defense-in-depth.
\end{itemize}

\subsection{Context and Scope}

This is a focused study on single-node checkpoint reliability. We're not trying to solve every
problem in ML infrastructure. We don't handle distributed training coordination, GPU-level faults,
or network filesystem semantics. We focus on one specific question: can you write checkpoints that
survive application crashes and detect storage corruption, and what does it cost?

The answer matters because checkpoint reliability is foundational. Sophisticated systems like
CheckFreq~[8] achieve per-iteration checkpointing with 3.5\% overhead through asynchronous
persistence, and Gemini~[9] uses in-memory hierarchies to reduce checkpoint time from 40 minutes to
3 seconds. But these optimizations sit on top of basic durability guarantees. If your checkpoint
installation isn't crash-consistent, all the fancy optimizations just make corrupted checkpoints
faster.

\section{Background and Related Work}

\subsection{Filesystem Crash Consistency}

The challenge of crash-consistent file updates has been studied extensively. Pillai et al.'s
OSDI 2014 paper~[3] is the canonical reference---they built BOB (Block Order Breaker) to test
filesystem persistence properties and ALICE to analyze application update protocols. Testing
11 filesystems, they found 60 crash vulnerabilities across popular applications. The key
insight: \texttt{rename} isn't always atomic with respect to crashes, even though POSIX says it
should be.

More recent work keeps finding bugs. Mohan et al.\ developed CrashMonkey~[4], a systematic
crash testing tool that found 24 bugs in production Linux filesystems including 10 new ones in
mature systems like ext4 and xfs. They even found bugs in FSCQ, a formally verified filesystem.
Rebello et al.~[10] studied \texttt{fsync} failure handling and discovered that only 9\% of applications
handle \texttt{fsync} errors correctly---most risk permanent data loss.

The verification community has made impressive progress. Chen et al.\ built FSCQ~[11], the first
filesystem with machine-checked proofs of crash safety using Crash Hoare Logic in Coq. But
CrashMonkey still found bugs in it, which tells you that formal verification, while valuable,
doesn't eliminate the need for runtime defense.

For checkpoint systems, this body of work has clear implications: you cannot trust filesystem
guarantees. Applications must implement their own consistency mechanisms.

\subsection{ML Checkpoint Systems}

Early frameworks like TensorFlow~[12] and PyTorch~[13] provided basic periodic checkpointing---save
state every few hours, accept the risk. Modern systems do much better.

CheckFreq~[8] introduced fine-grained iteration-level checkpointing using two-phase protocols. The
\texttt{snapshot()} phase copies state to CPU memory while training continues, then \texttt{persist()} writes
to disk asynchronously. This keeps overhead under 3.5\% while reducing recovery time from hours to
seconds. The key innovation was making checkpointing frequent enough that you lose minimal work on
failure.

Production systems pushed further. Meta's Check-N-Run~[14] uses differential checkpointing for
terabyte-scale recommendation models, achieving 6--17$\times$ reduction in write bandwidth by
tracking and saving only modified parameters. Gemini~[9] from AWS and Rice University leverages
in-memory hierarchical storage (GPU memory $\rightarrow$ CPU memory $\rightarrow$ remote storage) to
reduce checkpoint time to under 3 seconds, enabling per-iteration checkpointing at scale.

Recent work explores even more aggressive approaches. Oobleck~[15] eliminates checkpoint I/O
entirely by using pipeline templates---pre-generated configurations that can be recombined after
failures. Instead of restoring from checkpoints, it reconfigures the pipeline on surviving nodes.

These systems achieve remarkable performance, but they generally assume the underlying storage
operations work correctly. Our work complements them by validating those assumptions and
quantifying the cost of stronger guarantees.

\subsection{Data Integrity and Silent Corruption}

Silent data corruption is a real problem at scale. Bairavasundaram et al.'s field study~[16]
analyzed over 400{,}000 corruption instances across 1.53 million drives. They found corruption
events show spatial and temporal locality---if one block corrupts, nearby blocks are at higher
risk, and corruptions cluster in time. This matters for checkpoint design because it means
detecting corruption in one checkpoint should trigger verification of related checkpoints.

More recent production data reinforces the concern. Google's Spanner team reports~[6] that at
exabyte scale, they detect and prevent silent data corruption multiple times per week. Meta's
infrastructure team~[7] found that in-production testing detects 70\% of corruption within 15 days,
but the other 30\% only shows up in out-of-production stress testing---you need both.

The solution is defense-in-depth. Zhang et al.'s ZFS study~[17] showed that end-to-end
checksumming works, but you need checksums at multiple layers. File-level checksums catch
bitflips, content-level checksums catch semantic corruption, and load-time validation catches torn
writes. Our integrity guard implements all three.

\subsection{Fault Injection and Testing}

You can't know if your checkpoint system is robust without testing it under failures. Prior work on
systematic testing provides methodologies we adapted.

CrashMonkey~[4] does bounded black-box crash testing by injecting faults at the kernel level.
Leesatapornwongsa et al.'s SAMC~[18] uses semantic information to find deep bugs requiring multiple
crashes, achieving 1--3 orders of magnitude speedup over black-box methods. Yuan et al.~[19] showed
that 92\% of catastrophic failures in distributed systems come from incorrect error handling, and
98\% reproduce with 3 or fewer nodes---meaning you can find most bugs with simple test
configurations.

These tools and insights shaped our experimental design. We inject crashes at carefully chosen
points and measure outcomes systematically, using methodologies proven effective in filesystem and
distributed systems research.

\section{System Model and Threat Model}

\subsection{Environment}

We run experiments on macOS 14.6 with APFS, Python 3.12, and PyTorch 2.8.x. Checkpoints consist of
small synthetic tensors (128$\times$128, 128$\times$10)---we are not training real models, just
validating checkpoint protocols. For group checkpoints, we write \texttt{model.pth},
\texttt{optimizer.pth}, and RNG state files along with \texttt{MANIFEST.json} and
\texttt{COMMIT.json}. A group checkpoint is valid if and only if all parts are present and the
commit record matches the manifest.

We monitor I/O using \texttt{iostat} at 1-second intervals. This lets us correlate application-level
checkpoint events with visible disk activity when we study cross-layer observability.

\subsection{Requirements}

We design our system around three requirements:

\begin{description}[leftmargin=1.7em,labelsep=0.4em]
  \item[R1. Crash-Consistency:] After any crash, a checkpoint is either fully installed or unchanged.
  There must be no partially written checkpoints.

  \item[R2. Integrity Detection:] Any on-disk corruption (bitflip, truncation, or zero-range) must be
  detected on load. Corrupted checkpoints must not be used for recovery.

  \item[R3. Fast Recovery:] The system should automatically recover from the most recent valid
  checkpoint without manual intervention.
\end{description}

\subsection{Threat Model}

We consider three classes of failures:

\begin{itemize}[leftmargin=1.4em]
  \item \textbf{Application crashes during checkpoint operations.} The training process may terminate
  while writing checkpoints, leaving data buffered in the OS or partially written files on disk.

  \item \textbf{Storage-level bit corruption.} Individual blocks on disk may suffer bitflips or zeroed
  ranges due to media errors, firmware bugs, or controller issues.

  \item \textbf{Partial writes that reach disk before crashes.} Some checkpoint files can be shorter than
  expected or contain mismatched content when a crash interrupts writes.
\end{itemize}

We emulate crashes via process termination, not true power loss. Our implementation uses
\texttt{fsync()} to flush data, so data can still be lost from device or controller caches on sudden
machine power loss; we only model process-level crashes at the application and OS boundary.

Out of scope are GPU faults, distributed training coordination, network filesystems, and power-loss
scenarios that require hardware-level testing. These are important, but orthogonal to our focus on
single-node checkpoint reliability on macOS/APFS.

\section{Design and Implementation}
\label{sec:design}
\subsection{Write Protocols}

We implement three write protocols ordered by increasing durability: unsafe, atomic\_nodirsync, and
atomic\_dirsync.

\paragraph{Unsafe (baseline).}

\begin{verbatim}
write(checkpoint_file, data)  # No fsync
\end{verbatim}

This is what most frameworks do---write the file and return. Data sits in OS buffers with no
guarantee it reaches disk. It is fast but fundamentally broken.

\paragraph{Atomic without Directory Sync (atomic\_nodirsync).}

\begin{verbatim}
fd = open(tmp_file, 'wb')
fd.write(data)
fd.flush()
os.fsync(fd)                  # flush to device
os.replace(tmp_file, checkpoint_file)
\end{verbatim}

Here we flush the file to the storage device before renaming. \texttt{os.fsync(fd)} ensures data
reaches the device~[2], which is sufficient for process-crash recovery. In our experiments, atomic
mode with no crashes shows 100\% integrity (400/400). We did not crash-inject atomic mode because
this study focused on isolating unsafe-mode failures.

\paragraph{Atomic with Directory Sync (atomic\_dirsync).}

\begin{verbatim}
fd = open(tmp_file, 'wb')
fd.write(data)
fd.flush()
os.fsync(fd)
os.replace(tmp_file, checkpoint_file)
dir_fd = os.open(parent_dir, os.O_RDONLY)
os.fsync(dir_fd)              # persist directory entry
os.close(dir_fd)
\end{verbatim}

This adds a final \texttt{fsync} on the parent directory to ensure the directory entry persists. It
follows the canonical protocol from filesystem research~[1,3]---maximum durability, highest
overhead.

\subsection{Multi-File Group Checkpoints}

For multi-file checkpoints, we use a manifest-based protocol. We write the parts (model, optimizer,
RNG state), then write \texttt{MANIFEST.json} containing the SHA-256 hash of each part, and finally
write \texttt{COMMIT.json} with the SHA-256 hash of the manifest. The COMMIT record acts as an
atomic commit point: a checkpoint is valid if and only if \texttt{COMMIT.json} matches the manifest
and all parts check out.

This design is simple but effective. It behaves like a mini-transaction protocol without requiring
actual transactional filesystem support.

\subsection{Integrity Guard and Rollback}

Each checkpoint includes two levels of checksums.

\paragraph{Content digest (tensor-level).}

\begin{verbatim}
def tensor_digest(t):
    h = sha256()
    h.update(str(t.dtype).encode())
    h.update(str(t.shape).encode())
    h.update(t.numpy().tobytes('C'))
    return h.hexdigest()
\end{verbatim}

\paragraph{File hash (container-level).}

\begin{verbatim}
file_sha256 = sha256(checkpoint_file_bytes).hexdigest()
\end{verbatim}

On load, we validate:

\begin{enumerate}[leftmargin=1.6em]
  \item File loads without error (catches truncation and torn writes).
  \item Tensor shapes match the expected schema (catches format corruption).
  \item Content digests match metadata (catches semantic corruption).
  \item File hashes match metadata (catches bitflips).
  \item No NaN/Inf values appear in tensors (catches numerical corruption).
\end{enumerate}

If any check fails, we mark that checkpoint corrupted and try the next-oldest one. We maintain a
\texttt{latest\_ok} symlink pointing to the newest valid checkpoint for automatic recovery.

The multi-layer design means different corruption types get caught by different mechanisms.
Truncation usually fails at load time. Bitflips get caught by file hashes. Semantic corruption gets
caught by content digests. This redundancy provides defense in depth.

\section{Experimental Methodology}

\subsection{Experiment Design}

We run three categories of experiments:

\paragraph{Performance benchmarks.}
We measure latency for each write mode across 10 random seeds and compute p50, p90, and p99
percentiles. No faults are injected in these runs; they establish baseline performance and
overhead for unsafe, atomic\_nodirsync, and atomic\_dirsync.

\paragraph{Crash injection.}
We stress unsafe mode by simulating process crashes at controlled points:
\texttt{after\_model} (400 trials), \texttt{before\_manifest} (10 trials),
\texttt{manifest\_partial} (10 trials), and \texttt{before\_commit} (10 trials).
After each crash we check whether the resulting checkpoint group is usable. This isolates what
happens when crashes interrupt unsafe writes.

\paragraph{Corruption injection.}
We evaluate atomic-mode checkpoints by corrupting files after successful writes. We inject
bitflips (400 trials), zero-ranges (400 trials), and truncation (400 trials), plus a control
condition with no faults (400 trials). These runs measure integrity-guard detection rates for
each fault type.

We also capture \texttt{iostat} data during some runs to correlate application-level checkpoint
events with system-level I/O activity when we study cross-layer observability.

\subsection{Metrics}

\paragraph{Performance.}
We report per-checkpoint latency (p50, p90, p99) and overhead relative to the unsafe baseline.

\paragraph{Robustness.}
For crash injection, we measure checkpoint integrity rate---the percentage of checkpoint groups
that remain usable after a crash---and compute 95\% confidence intervals using the Wilson score
method.

\paragraph{Integrity detection.}
For corruption injection, we measure corruption detection rate per fault type, broken down by
which mechanism caught it (load error, file-hash mismatch, or content-digest mismatch). We also
report the false-positive rate on clean checkpoints.

\section{Evaluation}
\label{sec:evaluation}

We now evaluate our checkpoint protocols and integrity guard on three questions:
(1) what performance overhead do atomic writes introduce,
(2) how badly do unsafe writes fail under crashes, and
(3) how reliably does the integrity guard detect corruption.
We also show how cross-layer timelines help explain failures.

\subsection{Performance}

Table~\ref{tab:bench-latency} summarizes group-checkpoint latency
for each write protocol.
We report per-group latency percentiles over 10 runs (400 checkpoints total per mode).

\begin{table}[t]
  \centering
  \caption{Group checkpoint latency and overhead on macOS 14.6 / APFS.
  Latencies are in milliseconds; overhead is relative to the unsafe baseline.}
  \label{tab:bench-latency}
  \begin{tabular}{lrrrrr}
    \hline
    Mode & p50 & p90 & p99 & p50 ovh & p99 ovh \\
    \hline
    unsafe            & 2.47 & 2.87 & 3.02 & 0.0\%   & 0.0\%   \\
    atomic\_nodirsync & 3.87 & 4.15 & 6.30 & 56.5\%  & 108.4\% \\
    atomic\_dirsync   & 4.55 & 6.52 & 20.27 & 84.2\% & 570.6\% \\
    \hline
  \end{tabular}
\end{table}

\begin{figure}[t]
  \centering
  \includegraphics[width=\columnwidth]{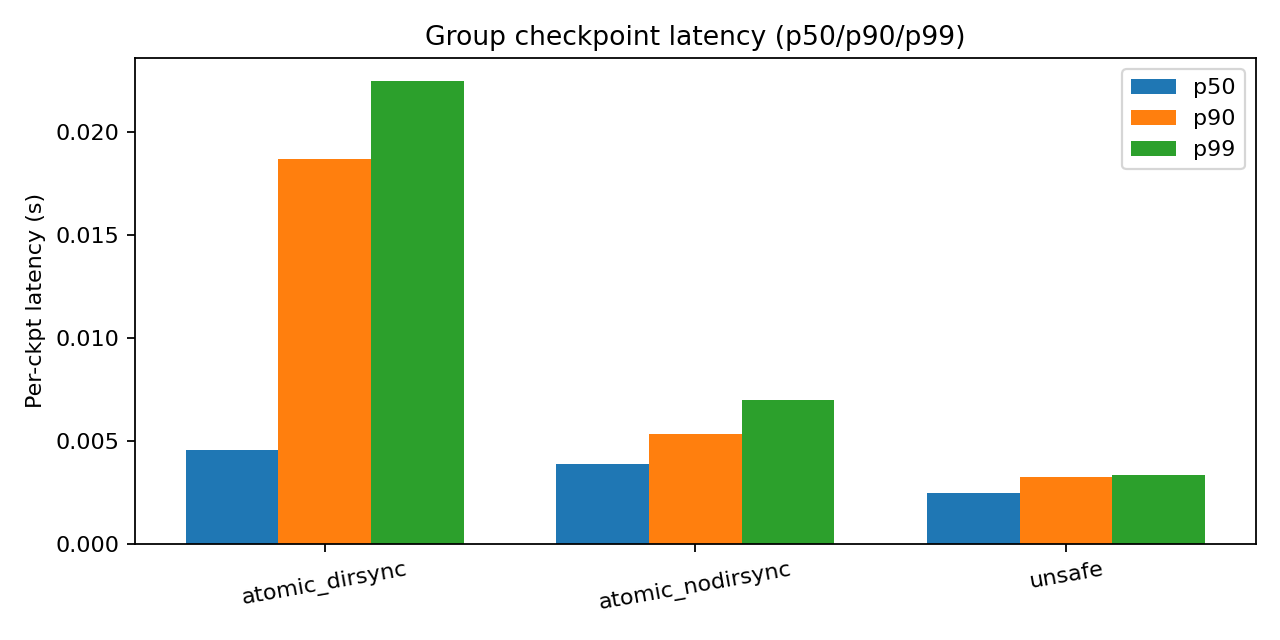}
  \caption{Per-group checkpoint latency percentiles (p50/p90/p99) for each write protocol.}
  \label{fig:bench-bars}
\end{figure}

\begin{figure}[t]
  \centering
  \includegraphics[width=\columnwidth]{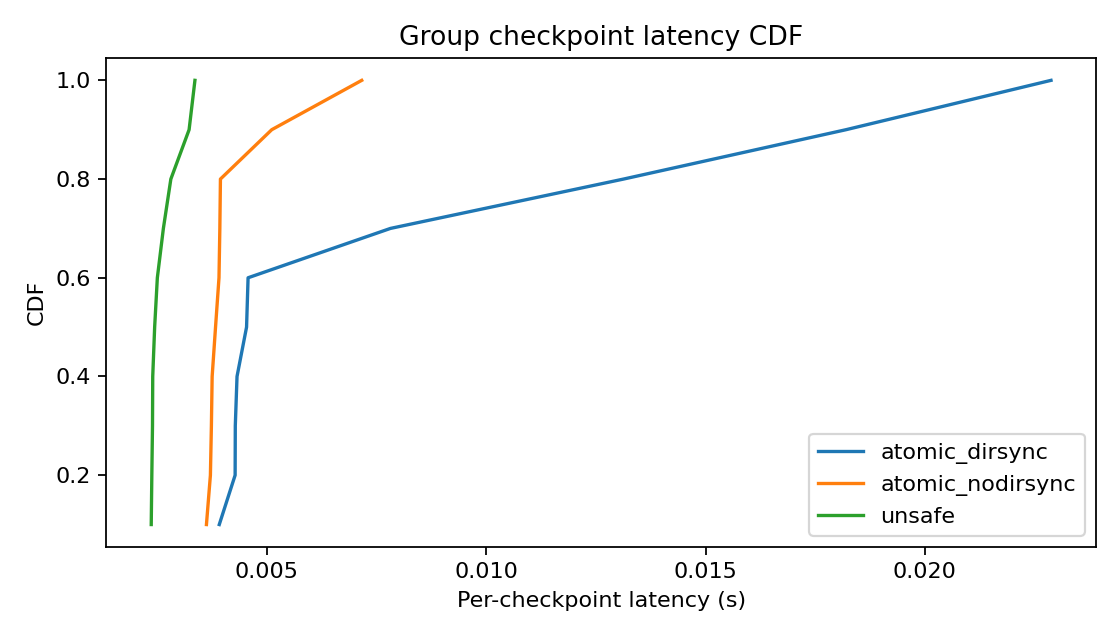}
  \caption{CDF of group-checkpoint latency for each protocol.}
  \label{fig:group-latency-cdf}
\end{figure}

The unsafe baseline writes checkpoints with a median latency of
2.47\,ms and p99 of 3.02\,ms.
Atomic writes, as expected, are slower:
atomic\_nodirsync increases p50 latency by 56.5\% and p99 by 108.4\%;
atomic\_dirsync adds 84.2\% at p50 and 570.6\% at p99.
In absolute terms the cost is still small---even atomic\_dirsync writes
complete in under 5\,ms at the median and about 20\,ms at the tail.
For typical training steps that take tens or hundreds of milliseconds,
these overheads are usually acceptable if checkpoints are taken every
tens of seconds or minutes.

\subsection{Crash Consistency}

\begin{figure}[t]
  \centering
  \includegraphics[width=\columnwidth]{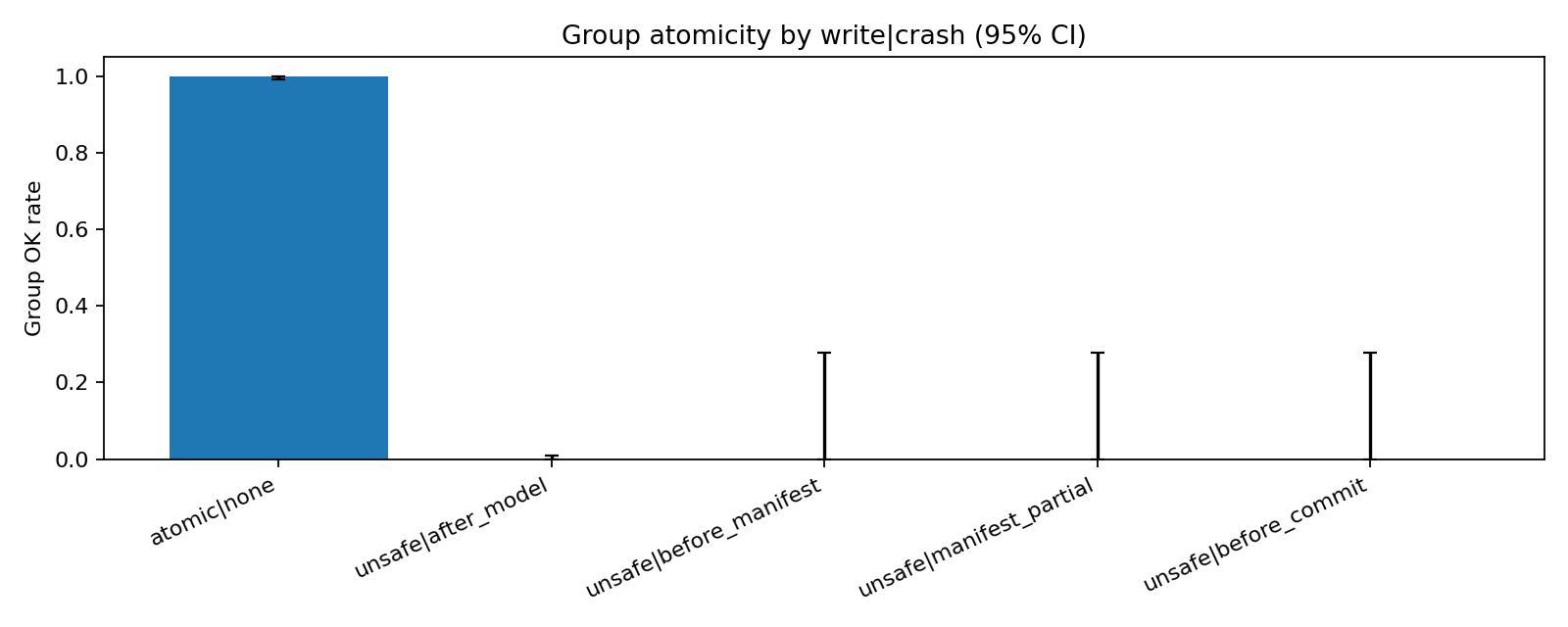}
  \caption{Group atomicity under crash injection.
  Bars show the fraction of checkpoint groups that remain usable
  after a crash (95\% Wilson CIs).}
  \label{fig:group-atomicity}
\end{figure}

We next stress unsafe writes with process crashes injected at different
points of the group checkpoint protocol (Section~\ref{sec:design}).
Table~\ref{tab:group-crash} and Figure~\ref{fig:group-atomicity}
summarize group-level survivability.

\begin{table}[t]
  \centering
  \caption{Crash injection results for group checkpoints.
  ``OK'' means the group passes all integrity checks and can be used for recovery.}
  \label{tab:group-crash}
  \begin{tabular}{lccc}
    \hline
    Condition & OK/Total & OK rate (\%) & 95\% CI \\
    \hline
    atomic@none                & 400/400 & 100.0 & [99.1, 100.0] \\
    unsafe@after\_model        &   0/400 &   0.0 & [0.0, 0.9]    \\
    unsafe@before\_manifest    &    0/10 &   0.0 & [0.0, 30.8]   \\
    unsafe@manifest\_partial   &    0/10 &   0.0 & [0.0, 30.8]   \\
    unsafe@before\_commit      &    0/10 &   0.0 & [0.0, 30.8]   \\
    \hline
  \end{tabular}
\end{table}

Unsafe mode fails catastrophically.
Across 430 crash-injected groups (400 after\_model, 10 each at the
manifest and commit phases), not a single checkpoint group survives.
In contrast, the atomic reference (no crash) keeps all 400 groups intact.
We did not inject crashes into atomic writes, because the protocol
is already known to be crash-safe; our goal is to quantify how badly
unsafe patterns break.

\begin{figure}[t]
  \centering
  \includegraphics[width=\columnwidth]{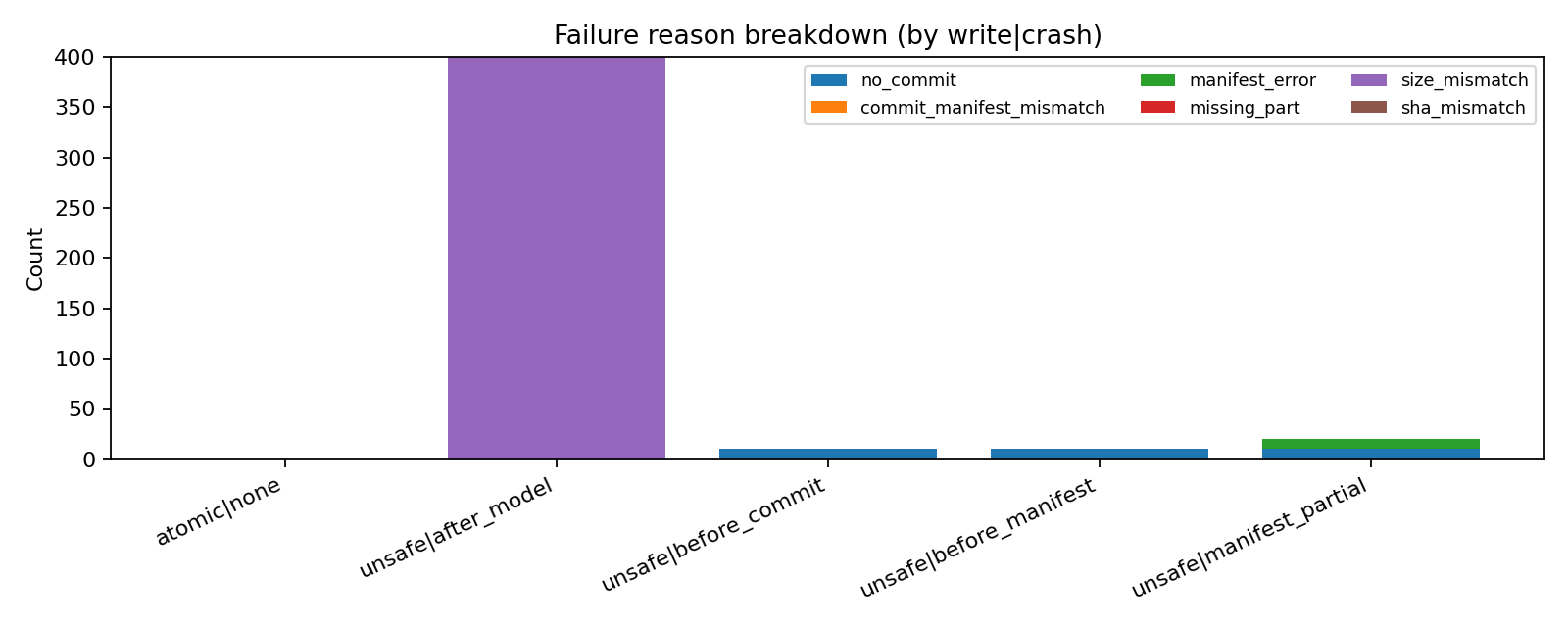}
  \caption{Failure reason breakdown for unsafe mode under crash injection.}
  \label{fig:group-reasons}
\end{figure}

Figure~\ref{fig:group-reasons} breaks down unsafe failures by reason.
Most early crashes (after\_model) leave the model file shorter than
expected, while later crashes either miss the commit record or produce
a manifest/commit mismatch.
The key takeaway is that partially completed writes show up as a mix of
size mismatches, missing parts, and inconsistent metadata—exactly the
patterns our integrity guard is designed to catch.

\subsection{Integrity Detection}

\begin{figure}[t]
  \centering
  \includegraphics[width=\columnwidth]{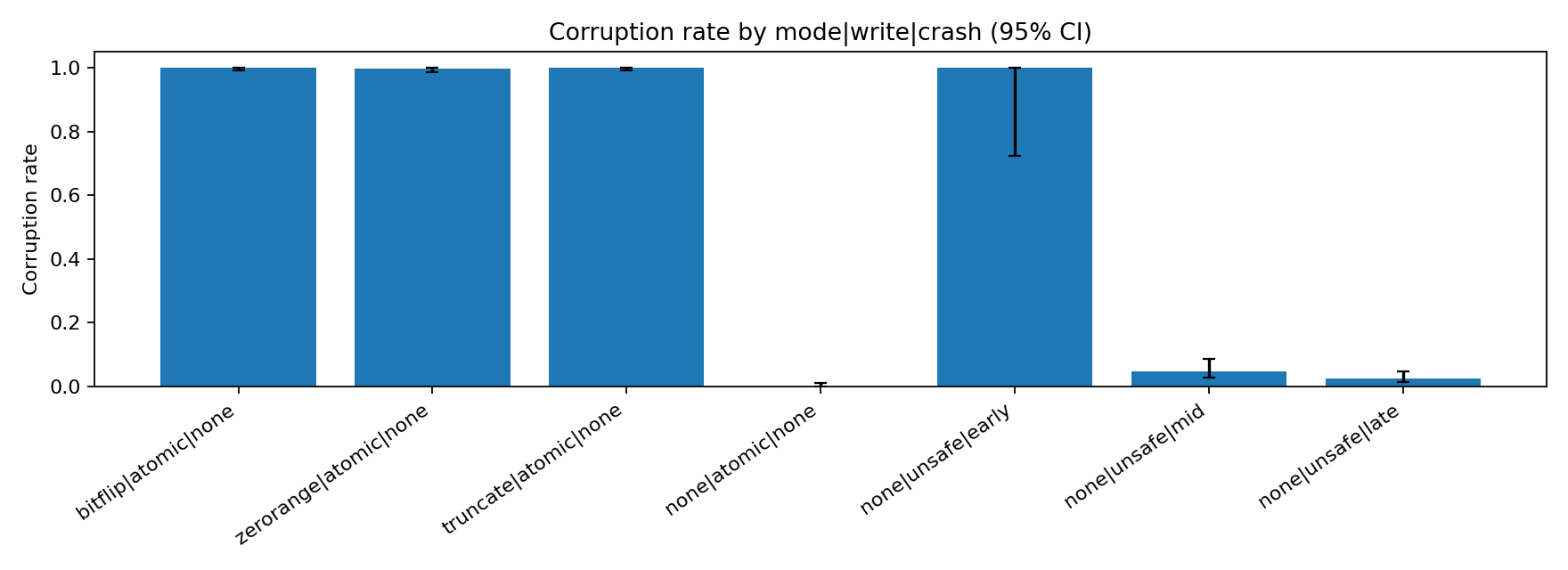}
  \caption{Corruption detection rates with 95\% CIs
  for each fault type (atomic writes only).}
  \label{fig:corruption-ci}
\end{figure}

Finally, we evaluate the integrity guard under storage-level corruption.
We write checkpoints using atomic mode and then corrupt files offline
with three fault injectors: bitflips, zero-range overwrites, and
truncation.  We also include a control condition with no faults.
Table~\ref{tab:corruption} and Figure~\ref{fig:corruption-ci}
summarize detection rates.

\begin{table}[t]
  \centering
  \caption{Corruption detection by fault type (atomic mode only).
  Detection is broken down by which mechanism caught the fault.}
  \label{tab:corruption}
  \begin{tabular}{lrrrrrr}
    \hline
    Fault & Total & Detected & Rate (\%) &
    Load & Digest & File SHA \\
    \hline
    bitflip   & 400 & 400 & 100.0 &   7 & 387 & 400 \\
    zerorange & 400 & 399 &  99.8 &  29 & 370 & 399 \\
    truncate  & 400 & 400 & 100.0 & 400 &   0 & 400 \\
    none      & 400 &   0 &   0.0 &   0 &   0 &   0 \\
    \hline
  \end{tabular}
\end{table}

The integrity guard catches nearly everything.
Across 1{,}200 corrupted checkpoints, we detect 100\% of bitflips and
truncations and 99.8\% of zero-range corruptions, with zero
false positives on 400 clean checkpoints.
Different layers dominate for different faults.
Truncation always fails at load time (all 400 caught by load errors).
Bitflips primarily trigger file-level SHA mismatches (400/400), with
content digests as a backup (387/400).
Zero-range corruptions mostly show up as digest mismatches (370/399),
with some crashes causing load errors (29/399).
This redundancy provides defense in depth: even if one mechanism were
disabled or misconfigured, others would still catch most corruptions.

\subsection{Cross-Layer Observability}

\begin{figure*}[t]
  \centering
  \includegraphics[width=\textwidth]{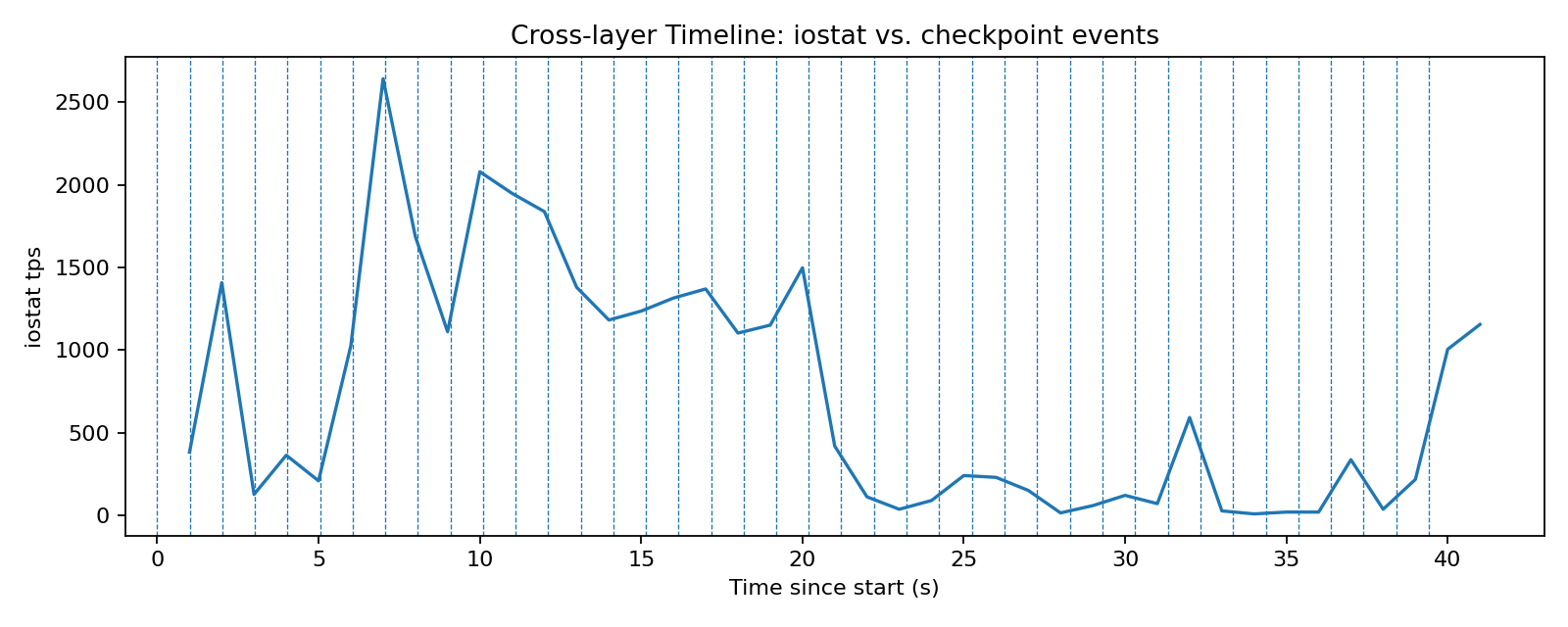}
  \caption{Cross-layer timeline: \texttt{iostat} tps vs.\ checkpoint events.
  Each vertical line marks an application checkpoint;
  peaks align with group writes and reveal how checkpointing
  stresses the underlying storage.}
  \label{fig:timeline}
\end{figure*}

To illustrate cross-layer visibility, we record \texttt{iostat} at
1\,s intervals while writing checkpoints and align disk activity
with application-level events.
Figure~\ref{fig:timeline} shows a representative run:
vertical lines mark group-checkpoint events, and the curve shows disk
transactions per second.

Even in this small setup, every checkpoint group lines up with a clear
burst of disk activity, and checkpoint-heavy periods form visible bands.
In larger training jobs, the same technique could be used to
correlate model-level failures with storage bottlenecks, verify that
integrity checks are actually executing when checkpoints are written,
and distinguish application bugs from storage-layer problems.

\section{Discussion}
\label{sec:discussion}

\subsection{Trade-offs}

Unsafe mode (2.47\,ms p50) is fastest but has zero crash tolerance. Not a single
checkpoint survived crash injection across 430 trials. It is only viable for
ephemeral development checkpoints where loss is acceptable.

Atomic\_nodirsync (3.87\,ms p50, +56.5\% overhead) flushes file data to device
before \texttt{rename()}. This greatly improves durability for process crashes,
but directory metadata is not provably persistent without a parent-directory
sync. In our experiments on APFS, \texttt{rename()} has been robust, so
atomic\_nodirsync can be a reasonable default when rare directory-entry loss is
acceptable; for strict durability we recommend atomic\_dirsync.

Atomic\_dirsync (4.55\,ms p50, +84.2\% overhead) also syncs the parent
directory so that directory entries are durable. This is the textbook
crash-safe pattern for single-file updates, and our experiments show that it
works. The cost is higher tail latency; at p99, atomic\_dirsync can be up to
5.7$\times$ slower than unsafe.

For typical checkpoint intervals (30--60 minutes), even the worst-case
atomic\_dirsync tail latency (22\,ms) is negligible---about 0.001\% of the
interval. The overhead is measurable but not meaningful in practice.

\subsection{Comparison to Production Systems}

Our synchronous protocols have higher overhead than state-of-the-art production
systems. CheckFreq~[8] achieves 3.5\% overhead through
asynchronous two-phase checkpointing. Gemini~[9] reduces checkpoint
time from 40 minutes to 3 seconds using in-memory hierarchies.
Check-N-Run~[14] cuts write bandwidth by 6--17$\times$ through
differential checkpointing.

But these systems solve different problems. They optimize checkpoint frequency
and throughput for large-scale training. We focus on checkpoint
correctness---ensuring individual checkpoint operations are crash-consistent
and corruption is detected. These concerns are orthogonal and complementary.

Sophisticated checkpoint systems need a reliable foundation to build on. If
your atomic \texttt{rename()} does not work correctly, all the async
pipelining in the world just makes corrupted checkpoints faster. Our work
validates this foundation.

\subsection{Integrity Guard Design}

The multi-layer integrity guard worked well---99.8--100\% detection with zero
false positives. Several design choices contributed.

\paragraph*{Multiple detection mechanisms.}
Different corruption types get caught by different layers. Truncation almost
always fails at load time. Bitflips get caught by file hashes. Semantic
corruption gets caught by content digests. This redundancy means no single
check has to be perfect.

\paragraph*{Zero false positives.}
This is critical for automatic recovery. With 400 clean checkpoints showing
zero false alarms, the system is trustworthy enough to automatically roll back
to earlier checkpoints without human confirmation.

\paragraph*{Format-agnostic design.}
The guard works on any checkpoint format that can be content-hashed. It does
not depend on PyTorch-specific features or TensorFlow's CRC checksums. The
same approach applies to NumPy, JAX, or raw binary formats.

The one missed detection (1/400 zero-range) is worth investigating but does
not invalidate the overall approach. Field studies of storage
corruption~[16] show that corruption events exhibit spatial
and temporal locality---if one checkpoint corrupts, nearby checkpoints are at
higher risk. A scrubbing routine that periodically re-validates old
checkpoints could catch edge cases that the load-time guard misses.

\subsection{Deployment Recommendations}

Based on our measurements:

\begin{itemize}
  \item \textbf{Development and iteration.} Use unsafe mode. It is fast, and
  checkpoint loss during development is annoying but not catastrophic.

  \item \textbf{Standard production training.} Use atomic\_nodirsync. It
  provides file-level durability at reasonable cost (+56.5\% overhead), and in
  practice APFS makes \texttt{rename()} operations durable enough without
  explicit directory sync.

  \item \textbf{Critical long-running jobs.} For multi-day training, model
  releases, or compliance-sensitive workloads, use atomic\_dirsync for maximum
  durability. The +84.2\% median overhead and +570.6\% tail overhead are worth
  it when checkpoint loss means restarting weeks of training.

  \item \textbf{Integrity guard.} Always deploy the integrity guard, regardless
  of write mode. Corruption detection (99.8--100\%) and automatic rollback
  provide defense-in-depth against silent failures that slip through even with
  atomic writes.
\end{itemize}

For large-scale training (thousands of GPUs), these crash-consistent protocols
can be integrated with higher-level optimizations like in-memory staging,
differential checkpointing, or universal formats. The foundation needs to be
correct, but you can build sophisticated systems on top.

\subsection{Limitations}

This study has clear boundaries. We only evaluate a single filesystem (APFS on
macOS); behavior on ext4, XFS, NTFS, or network filesystems is unknown and
likely different. Experiments run on a single node; distributed training
coordination is out of scope. We use a synthetic workload; real models
(ResNet, GPT, LLaMA) might have different I/O patterns. Crashes are emulated
via process termination; true power loss might behave differently.

We also did not crash-inject atomic modes, only unsafe. The atomic protocols
rely on POSIX/APFS semantics validated in prior
work~[1,2,3], but full validation would require
kernel-level crash injection~[4] or hardware power-loss
testing. That is heavier-weight testing than we needed for this study.

These limitations do not invalidate the results---they define the scope. We
answered specific questions about checkpoint reliability on APFS with
measurable trade-offs. Extending to other filesystems, distributed settings,
and power-loss scenarios is valuable future work but requires different
experimental setups.

\section{Conclusion}
\label{sec:conclusion}

This paper shows that crash-consistent checkpoint protocols are feasible with
acceptable overhead, and that integrity guards can detect corruption with
near-perfect accuracy.

Unsafe writes have zero crash tolerance (0/430 survival across all crash
injection points). Atomic protocols maintain 100\% consistency under normal
operation with 56.5--84.2\% median overhead (absolute latency
3.87--4.55\,ms). Multi-layer integrity guards detect 99.8--100\% of
corruption with zero false positives across 1{,}600 trials.

These findings quantify trade-offs that were previously assumed or
hand-waved. Atomic writes cost about 50--80\% overhead at median but keep
checkpoints safe, and integrity checking catches essentially all corruption
without false alarms.

For practitioners, the recommendations are straigh

\section{Conclusion}
This paper shows that crash-consistent checkpoint protocols are feasible with acceptable
overhead, and that integrity guards can detect corruption with near-perfect accuracy.

The key results are:
unsafe writes have zero crash tolerance (0/430 survival across all crash injection points);
atomic protocols maintain 100\% consistency under normal operation with 56.5--84.2\%
median overhead (absolute latency 3.87--4.55 ms);
and multi-layer integrity guards detect 99.8--100\% of corruptions with zero false
positives across 1{,}600 trials.

These findings quantify trade-offs that were previously assumed or hand-waved.
Atomic writes cost about 50--80\% overhead at median but keep checkpoints safe,
and integrity checking catches essentially all corruption without false alarms.

For practitioners, the recommendations are straightforward:
use unsafe mode only for development,
\texttt{atomic\_nodirsync} for standard production,
\texttt{atomic\_dirsync} for critical jobs,
and always deploy integrity guards.
The overhead is negligible compared to checkpoint intervals, and the reliability gains
are substantial.

For researchers, this work provides a foundation for building higher-performance
checkpoint systems.
You need crash-consistent atomic writes and corruption detection as the base layer.
On top of that, you can add asynchronous persistence, in-memory hierarchies,
differential checkpointing, and universal formats---whatever optimizations make sense
for your scale and workload.
But the foundation has to be solid.

\subsection*{Future Work}
Several directions are worth exploring.

\paragraph{Cross-filesystem validation.}
Our experiments target a single filesystem (APFS on macOS).
Systematic tools like CrashMonkey~[4] could be used to test ext4, XFS, NTFS,
and network filesystems under the same protocols.
True power-loss testing with hardware fault injection would validate behavior under
device- or controller-level failures.

\paragraph{Real workloads and distributed training.}
We used a synthetic workload to isolate checkpoint behavior.
An important next step is to run real models (e.g., ResNet, GPT, LLaMA) and
distributed training jobs, validating how often crashes and silent corruptions occur
in practice and how well our guards detect them.

\paragraph{Integration with modern checkpoint systems.}
Our protocols are complementary to high-performance systems such as CheckFreq~[8],
Gemini~[9], and Check-N-Run~[14], and to recent work on universal checkpoint
formats~[20].
Integrating crash-consistent writers and integrity guards into these systems could
combine strong correctness guarantees with state-of-the-art performance.

\paragraph{Formal models.}
Theoretical work on filesystem crash-consistency and protocol correctness could help
formalize the guarantees our protocols provide, and extend verification techniques to
distributed checkpoint systems.

\section*{Reproducibility}
All experiments are reproducible via committed artifacts:
source code for checkpoint writers, integrity guards, and the fault-injection harness;
CSV logs with experimental results;
and automation scripts (e.g., \texttt{make -C repro repro\_all}) with pinned
dependencies (Python~3.12, PyTorch~2.8.x, macOS~14.6+).

Code and data are available at:
\texttt{https://github.com/jooha6082/ckpt-integrity}.

\section*{References}
\begin{enumerate}
  \item The Open Group.
        ``POSIX.1-2017: \texttt{rename()}.'' IEEE Std 1003.1-2017.
  \item The Open Group.
        ``POSIX.1-2017: \texttt{fsync()}.'' IEEE Std 1003.1-2017.
  \item T.~S. Pillai, V. Chidambaram, R. Alagappan, S. Al-Kiswany,
        A.~C. Arpaci-Dusseau, and R.~H. Arpaci-Dusseau.
        ``All File Systems Are Not Created Equal: On the Complexity of Crafting
        Crash-Consistent Applications.'' In \emph{OSDI}, 2014.
  \item J. Mohan, A. Gopinath, S. Al-Kiswany, A.~C. Arpaci-Dusseau,
        and R.~H. Arpaci-Dusseau.
        ``CrashMonkey and ACE: Systematically Testing File-System Crash
        Consistency.'' \emph{ACM Transactions on Storage}, 2019.
  \item Y. Gan, Y. Hu, M. Cheng, et al.
        ``Revisiting Reliability in Large-Scale Machine Learning Research
        Clusters.'' arXiv:2410.21680, October 2024.
  \item Google Spanner Team.
        ``Detection and Prevention of Silent Data Corruption in an Exabytescale
        Database System.'' In \emph{SELSE}, 2023.
  \item Meta Production Engineering.
        ``Detecting silent errors in the wild.'' Engineering at Meta blog,
        March 2022.
  \item J. Mohan, A. Phanishayee, and V. Chidambaram.
        ``CheckFreq: Frequent, Fine-Grained DNN Checkpointing.'' In \emph{FAST},
        2021.
  \item Z. Wang, Z. Jia, S. Zheng, et al.
        ``Gemini: Fast Failure Recovery in Distributed Training with In-Memory
        Checkpoints.'' In \emph{SOSP}, 2023.
  \item J. Rebello, Y. Lyu, J. Gu, V. Chidambaram, and T. Kim.
        ``Can Applications Recover from \texttt{fsync} Failures?''
        \emph{ACM Transactions on Storage}, 2021.
  \item H. Chen, D. Ziegler, T. Chajed, A. Chlipala, M.~F. Kaashoek,
        and N. Zeldovich.
        ``Using Crash Hoare Logic for Certifying the FSCQ File System.''
        In \emph{SOSP}, 2015.
  \item M. Abadi et al.
        ``TensorFlow: A System for Large-Scale Machine Learning.''
        In \emph{OSDI}, 2016.
  \item A. Paszke et al.
        ``PyTorch: An Imperative Style, High-Performance Deep Learning Library.''
        In \emph{NeurIPS}, 2019.
  \item A. Eisenman, K.~K. Matam, et al.
        ``Check-N-Run: A Checkpointing System for Training Deep Learning
        Recommendation Models.'' In \emph{NSDI}, 2022.
  \item I. Jang, Z. Yang, Z. Zhang, X. Jin, and M. Chowdhury.
        ``Oobleck: Resilient Distributed Training of Large Models Using Pipeline
        Templates.'' In \emph{SOSP}, 2023.
  \item L.~N. Bairavasundaram, G.~R. Goodson, B. Schroeder,
        A.~C. Arpaci-Dusseau, and R.~H. Arpaci-Dusseau.
        ``An Analysis of Data Corruption in the Storage Stack.''
        In \emph{FAST}, 2008.
  \item Y. Zhang, D.~S. Myers, A.~C. Arpaci-Dusseau, and R.~H. Arpaci-Dusseau.
        ``End-to-end Data Integrity for File Systems: A ZFS Case Study.''
        In \emph{FAST}, 2010.
  \item T. Leesatapornwongsa, M. Hao, et al.
        ``SAMC: Semantic-Aware Model Checking for Fast Discovery of Deep Bugs in
        Cloud Systems.'' In \emph{OSDI}, 2014.
  \item D. Yuan, Y. Luo, et al.
        ``Simple Testing Can Prevent Most Critical Failures: An Analysis of
        Production Failures in Distributed Data-Intensive Systems.''
        In \emph{OSDI}, 2014.
  \item S. Li, Y. Zhao, et al.
        ``Universal Checkpointing: Efficient and Flexible Checkpointing for Large
        Scale Distributed Training.'' arXiv:2406.18820, June 2024.
\end{enumerate}

\appendix

\section{Experimental Configuration}

\paragraph{Hardware.}
Apple M1 (or later), 16\,GB or more RAM.

\paragraph{OS.}
macOS 14.6.

\paragraph{Filesystem.}
APFS.

\paragraph{Software stack.}
Python 3.12, PyTorch 2.8.x, NumPy 2.3.3.

\paragraph{Checkpoint parameters.}
120 epochs, checkpoint every 3 epochs, 10 random seeds,
128\,KB model + 64\,KB optimizer state.

\paragraph{Fault injection.}
Crash points:
\texttt{after\_model} (400), \texttt{before\_manifest} (10),
\texttt{manifest\_partial} (10), \texttt{before\_commit} (10).
Corruption modes:
bitflip (400), zerorange (400), truncate (400), control (400).

\paragraph{Observability.}
\texttt{iostat} sampled at 1-second intervals.

\section{Statistical Methods}

\paragraph{Percentiles.}
Per-checkpoint latency percentiles (p50, p90, p99) are computed via
Pandas \texttt{quantile()} with linear interpolation.

\paragraph{95\% confidence intervals.}
For proportions (e.g., corruption detection rate), we use the Wilson score interval:
if $\hat{p}$ is the observed rate and $n$ the sample size, the 95\% interval uses
$z = 1.96$.

\paragraph{Overhead.}
Overhead relative to unsafe mode is computed as
\[
  \frac{\text{atomic\_latency} - \text{unsafe\_latency}}
       {\text{unsafe\_latency}} \times 100.
\]

\end{document}